\documentclass[aps,pra,floatfix,twocolumn,groupedaddress,superscriptaddress,footinbib,notitlepage,showpacs]{revtex4-1}
\usepackage{graphicx,graphics,times,bm,bbm,bbold,amssymb,amsmath,amsfonts,dsfont,hyperref,color}
\hypersetup{colorlinks,linkcolor=blue,citecolor=blue,urlcolor=blue}
\let\oldsqrt\sqrt
\def\sqrt{\mathpalette\DHLhksqrt}
\def\DHLhksqrt#1#2{%
\setbox0=\hbox{$#1\oldsqrt{#2\,}$}\dimen0=\ht0
\advance\dimen0-0.2\ht0
\setbox2=\hbox{\vrule height\ht0 depth -\dimen0}%
{\box0\lower0.4pt\box2}}

\DeclareFontFamily{OT1}{pzc}{}
\DeclareFontShape{OT1}{pzc}{m}{it}%
              {<-> s * [1.25] pzcmi7t}{}
\DeclareMathAlphabet{\mathpzc}{OT1}{pzc}%
                                 {m}{it}

\begin{document}

\title{Boundary-induced coherence in the staggered quantum walk on different topologies}

\author{J. Khatibi Moqadam}
\author{A. T. Rezakhani}
\affiliation{Department of Physics, Sharif University of Technology, Tehran 14588, Iran}
\date{\today}

\begin{abstract}
The staggered quantum walk is a type of discrete-time quantum walk model without a coin which can be generated on a graph using particular partitions of the graph nodes. 
We design Hamiltonians for potential realization of the staggered dynamics on a two-dimensional lattice composed of superconducting microwave resonators connected with tunable couplings.
The naive generalization of the one-dimensional staggered dynamics generates two uncoupled one-dimensional quantum walks thus more complex partitions need to be employed.
However, by analyzing the coherence of the dynamics, we show that the quantumness of the evolution corresponding to two independent one-dimensional quantum walks can be elevated to the level of a single two-dimensional quantum walk, only by modifying the boundary conditions.
In fact, by changing the lattice boundary conditions (or topology), we explore the walk on different surfaces such as torus, Klein bottle, real projective plane and sphere.
The coherence and the entropy reach different levels depending on the topology of the surface.
We observe that the entropy captures similar information as coherence, thus we use it to explore the effects of boundaries on the dynamics of the continuous-time quantum walk and the classical random walk.
\end{abstract}


\maketitle

\section{Introduction}
\label{sec:intr}

The quantization of the classical random walks can be achieved in different ways, among which the coined discrete-time quantum walk and the continuous-time quantum walk (CTQW) are widely known \cite{portugal2013quantum}. In particular, the \mbox{staggered} quantum walk (SQW) \citep{portugal2016graph}, a discrete-time quantum walk model without a coin, has attracted growing attention recently. This model has various interesting features, for example, it enables efficient quantum search algorithms in two dimensions---providing approximately quadratic speedup with respect to the classical case \cite{portugal2017quantum}. The two-dimensional (2D) CTQW model (including no coin either), however, fails to speed up the search algorithms, unless additional degrees of freedom as extra sites are embedded into the lattice \cite{chils2014spatial}. Interestingly, the SQW somehow has already taken into account this idea, where, in fact, the coin degrees of freedom has been converted to extra nodes in the corresponding lattice. Besides, in terms of physical implementations, the SQW is more favorable than the coined discrete-time quantum walk due to the very absence of the coin. Although removing the coin operator comes with the price of demanding a dynamical graph for the walk evolution, by employing the state of the art superconducting circuits technology, the dynamics can be realized in a lattice of superconducting microwave resonators with controllable couplings \cite{moqadam2017staggered}.

To generate the staggered dynamics on a graph, distinct partitions of the set of graph nodes are considered. The elements of each partition contains only the first-neighbor vertices, namely the nodes which are all connected by edges. Such partitions are called tessellations, and associated to each tessellation a unitary operator is constructed. The condition that the set of graph edges should be covered in the union of all tessellations determines the number of required unitary operators. The SQW evolution operator is obtained by multiplying all those operators. Being devised principally for constructing the staggered operators, the partitioning process can also be used to obtain the staggered Hamiltonians \cite{portugal2017staggered-hamiltonians}.

The connection between the SQW and other quantum walk models has been explored in Refs. \cite{portugal2016establishing, portugal2016staggered-qip, portugal2015one-dimensional, philipp2017exact}. Different discrete-time quantum walk models including the 2-tessellable SQW
can be analyzed under a common framework called the two-partition model \cite{konno2018partition}, inspired by the staggered dynamics.

The 1D coined quantum walk model was used to explore topological phases in condensed matter
systems \cite{kitagawa2010exploring,asboth2012symmetries,asboth2013bulk,obuse2015unveiling,cedzich2016bulk}.
Including a position-dependent phase shift in each step of the walk, the dynamics of a charged particle in
the presence of external fields can be simulated \cite{genske2013electric,cedzich2013propagation}.
Using such Bloch oscillating quantum walks, topological invariants corresponding to the split-steps quantum
walks \cite{kitagawa2010exploring} can be directly measured, employing a superconducting microwave resonator
cavity coupled to a transmon qubit \cite{ramasesh2017direct,flurin2017observing}.
The 1D SQW, on the other hand, is related the Su-Schrieffer-Heeger
model \cite{su1979soliton,asboth2016short} which was also used in simulating topological
invariants \cite{li2014topological,gu2017topological,meier2016observation}.

The effects of the boundary conditions determining the topological properties of the system are indispensable in the dynamics associated with topological insulators \cite{hasan2010topological,qi2011topological} and topological quantum computing \cite{roy2017topological,lahtinen2017short}.
The properties of discrete- and continuous-time quantum walks on different topologies have already been studied in Refs. \cite{ambainis2001one,konno2002limit,konno2002absorption,bach2004one,rohde2012entanglement, kendon2006quantum,konno2008quantum}. For example, the quadratic speed up of quantum-walk-based search algorithms was obtained on lattices with periodic boundary conditions, that is with a torus topology \cite{portugal2013quantum,chils2014spatial,portugal2017quantum}.

In this paper, we address the implementation of the 2D SQW referring to a lattice of superconducting microwave resonators coupled with adjustable devices. Two types of Hamiltonians are constructed, one of which generates two 1D SQWs and the other one generates a 2D SQW. The quantumness \cite{baumgratz2014quantifying,shahbeigi2018quantum} of those dynamics---quantified by the coherence of the walker during the evolution---are different. However, by analyzing the SQW dynamics on a lattice with different boundary conditions, resembling 2D manifolds such as torus, Klein bottle, real projective plane and sphere \cite{nakahara2003geometry}, we show that the coherence can be increased by using twisted boundary conditions. We also analyze the dynamics of the entropy, which appears to be qualitatively similar to the behavior of coherence in the dynamics; hence we use it to quantify the quantumness of the CTQW. The boundary-independence of the entropy associated with the classical random walk dynamics supports that the boundary-induced coherence in quantum walk dynamics can be reflected by the entropy.

The paper is organized as follows. In Sec. \ref{sec:2}, we describe how to construct 2D SQW Hamiltonians and to realize the corresponding dynamics. In Sec. \ref{sec:3}, we use different boundary conditions to change the topology of the lattice and find their corresponding Hamiltonians. The properties of the walk dynamics on these topologies are analyzed in Sec. \ref{sec:4}. In Secs.\ref{sec:5} and \ref{sec:6}, the properties of the CTQW and the classical random walk are explored. Summary and further discussions are given in Sec. \ref{sec:7}.

\section{Staggered Quantum walk dynamics}
\label{sec:2}

Consider a lattice of harmonic resonators whose dynamics is described by the time-dependent tight-binding Hamiltonian with
switchable couplings ($\hbar\equiv 1$),
\begin{equation}
\label{eq:tight-binding}
H(t) = \sum_{n} \omega_n  a^{\dagger}_n a_n -
                 \sum_{\langle n,m\rangle} \kappa_{nm}(t)
                 ( a^{\dagger}_{n} a_{m} + a^{\dagger}_{m} a_n ),
\end{equation}
where $\omega_n$ are the resonators frequencies, $a^{\dagger}_n$ and $a_n$ are the creation and annihilation operators satisfying $[a_n,a^{\dagger}_m]=\delta_{nm}$, and $\kappa_{nm}(t)$ are the switchable couplings between nearest-neighbor resonators (denoted by $\langle n,m\rangle$).
We restrict the system to the ``single-photon" regime $\sum_{n} \langle a^{\dagger}_n a_n \rangle=1$, where $\langle a^{\dagger}_n a_n\rangle$ is the average of the operator in the system state. In this regime, the canonical basis for the lattice Hilbert space is associated with the presence of photon at each single site (resonator) of the lattice.
The physical implementations of such Hamiltonian, with arrays of superconducting microwave resonators coupled through superconducting quantum interference devices (\textsc{squid}s), have already been investigated in Refs. \cite{moqadam2017staggered,peropadre2013tunable,baust2015tunable,wulschner2015tunable}. The tunable couplings in those systems are achieved by adjusting the magnetic flux threading the \textsc{squid} loop. The typical frequencies are of the order of GHz for the resonators and MHz for the couplings.

In the following, we describe how to adjust the couplings in the system Hamiltonian (\ref{eq:tight-binding}) to realize the SQW dynamics on a 2D lattice. The rigorous mathematical framework for constructing the SQW on a generic graph has been put forward in Refs. \cite{portugal2016staggered-qip,portugal2016graph} (and a SQW-based quantum search algorithms was presented in Ref. \cite{portugal2017quantum}). Starting with the 1D SQW implementation of Ref. \cite{moqadam2017staggered}, we give explicit forms for the 2D SQW Hamiltonians using the tight-binding model with tunable couplings.
The SQW includes at least two unitary operators which are obtained according to independent partitions of the graph vertices. The elements in each partition should contain only the neighboring vertices---those that are connected by some edges (a single vertex is also accepted). Whereas each partition contains all the graph vertices, it includes only part of the graph edges. New partitions are then considered to include the edges not covered. Partitioning the graph vertices is performed as many times as required so that all edges of the graph are covered in the union of the partitions. Note that the intersection of the partitions should contain no edges. Associated to each partition, a unitary operator is defined and the multiplication of all such unitaries is the SQW operator. In the extended version of the SQW \cite{portugal2017staggered-hamiltonians} the graph partitions are used to construct the SQW Hamiltonians which then generate the walk operators.

\begin{figure}[tp]
\includegraphics[scale=.35]{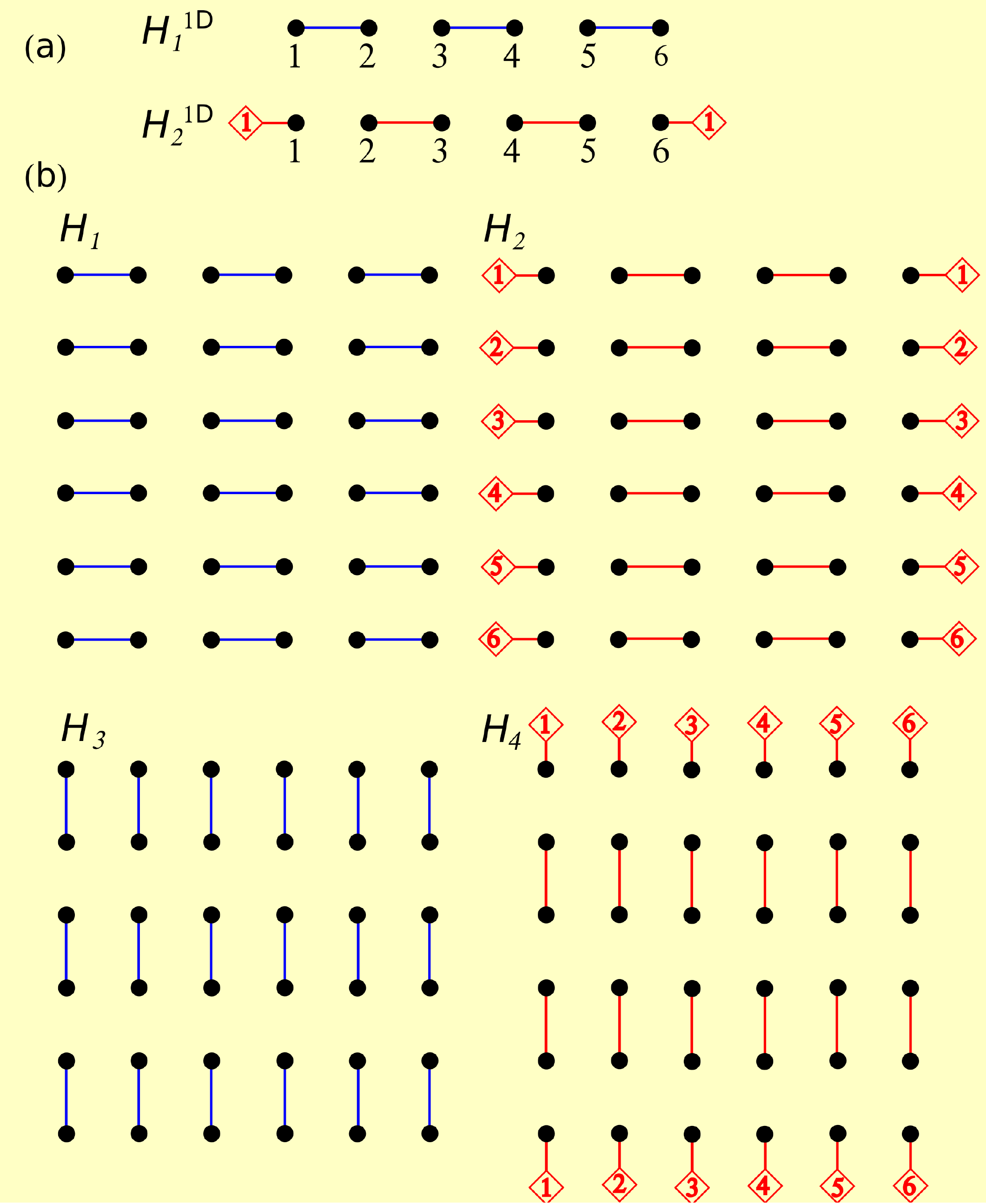}
\caption{Configuration of the turned-on couplings in the Hamiltonian (\ref{eq:tight-binding}) to realize the SQW Hamiltonians for the 1D lattice in
(a) and the 2D lattice in (b), with $d=6$. The ending sites specified with the same numbers inside the diamonds are connected. To generate each Hamiltonian all the sites are contributed. However, the set of all possible edges is divided into two subsets in the 1D case and four subsets in the 2D case.}
\label{fig:hamiltonians_naive}
\end{figure}

Realizing the SQW on a 1D lattice with $d$ sites requires two unitary operators corresponding to two time-independent Hamiltonians \cite{moqadam2017staggered}. Each Hamiltonian is comprised of a collection of disjoint pairs of interacting resonators. Turning on those couplings in the Hamiltonian (\ref{eq:tight-binding}) which couple odd-even resonators generates
\begin{equation}
\label{eq:H_1D1}
H^{\mathrm{1D}}_1 = \openone _{d/2} \otimes (\omega \openone _{2} - \kappa \sigma_x),
\end{equation}
where $\openone_{d/2}$ is the $(d/2)\times(d/2)$ identity operator with even $d$, $\sigma_{x}=\left(\begin{smallmatrix}0&1 \\1 & 0 \end{smallmatrix} \right)$ is the $x$-Pauli matrix, all the resonators are considered at resonance at frequency $\omega$, and all the nonzero couplings
are taken to be $\kappa$. Moreover, the interaction (hopping) term in the Hamiltonian (\ref{eq:tight-binding}) for any pair of resonators
is written in terms of $\sigma_x$. The Hamiltonian (\ref{eq:H_1D1}) describes the collection of disjoint odd-even pairs of interacting resonators.
The configuration of the turned-on couplings for Hamiltonian (\ref{eq:H_1D1}), with $d=6$, can be seen
in Fig. \ref{fig:hamiltonians_naive} (a).

In the same manner, the couplings between even-odd resonators in the Hamiltonian (\ref{eq:tight-binding}) can
be turned on to generate
\begin{equation}
\label{eq:H_1D2}
H^{\mathrm{1D}}_2 =
\begin{pmatrix}
 \omega & \mathbf{0} & -\kappa \\
 \mathbf{0} & \openone _{(d-2)/2} \otimes (\omega \openone _{2} - \kappa \sigma_x) & \mathbf{0} \\
 -\kappa & \mathbf{0} & \omega \\
\end{pmatrix},
\end{equation}
which describes a collection of disjoint even-odd pairs of resonators. The periodic boundary conditions
are used here. In Fig. \ref{fig:hamiltonians_naive} (a), the configuration of the turned-on couplings for
this case can also be seen. Note that the Hamiltonians (\ref{eq:H_1D1}) and (\ref{eq:H_1D2}) do not commute.

The SQW is implemented by repeatedly switching the Hamiltonian (\ref{eq:tight-binding}) between $H^{\mathrm{1D}}_1$ and $H^{\mathrm{1D}}_2$.
In fact, each step of the walk consists of generating $H^{\mathrm{1D}}_1$ in the time interval $[0,\tau)$ followed by generating $H^{\mathrm{1D}}_2$ in $[\tau,2\tau)$. More details for the walk on a 1D array can be found in Ref. \cite{moqadam2017staggered}.

For a 2D lattice, each step of the SQW is realized by applying four unitary operators corresponding to four time-independent Hamiltonians \cite{portugal2017quantum}. Suppose that the lattice has $N=d\times d$ sites with even $d$ and periodic boundary conditions. This lattice contains $2d^2$ edges (hence couplings). Again, each staggered Hamiltonian is constructed by turning on only parts of the couplings in the form (\ref{eq:tight-binding}) such that the lattice comprises a collection of disjoint pairs of interacting resonators. The pairs can be selected row-by-row or column-by-column (bearing in mind that only neighboring sites can be paired). The first SQW Hamiltonian can be constructed by switching on those couplings in the Hamiltonian form (\ref{eq:tight-binding}) that generate odd-even pairs in each row of the 2D lattice,
\begin{equation}
\label{eq:H_1_naive}
H_1 = \openone _d \otimes H^{\mathrm{1D}}_1.
\end{equation}
The second Hamiltonian is obtained by generating even-odd pairs in each row of the lattice,
\begin{equation}
\label{eq:H_2_naive}
H_2 = \openone _d \otimes H^{\mathrm{1D}}_2.
\end{equation}
Finally, two more Hamiltonians are constructed by creating pairs of interacting resonators in the columns
of the lattice,
\begin{align}
\label{eq:H_3_naive}
H_3 &= H^{\mathrm{1D}}_1 \otimes \openone _d,\\
\label{eq:H_4_naive}
H_4 &= H^{\mathrm{1D}}_2 \otimes \openone _d.
\end{align}
The configuration of the turned-on couplings for the Hamiltonians (\ref{eq:H_1_naive}) - (\ref{eq:H_4_naive}) (with $d=6$) are depicted in Fig. \ref{fig:hamiltonians_naive} (b).

\begin{figure}
\includegraphics[scale=.35]{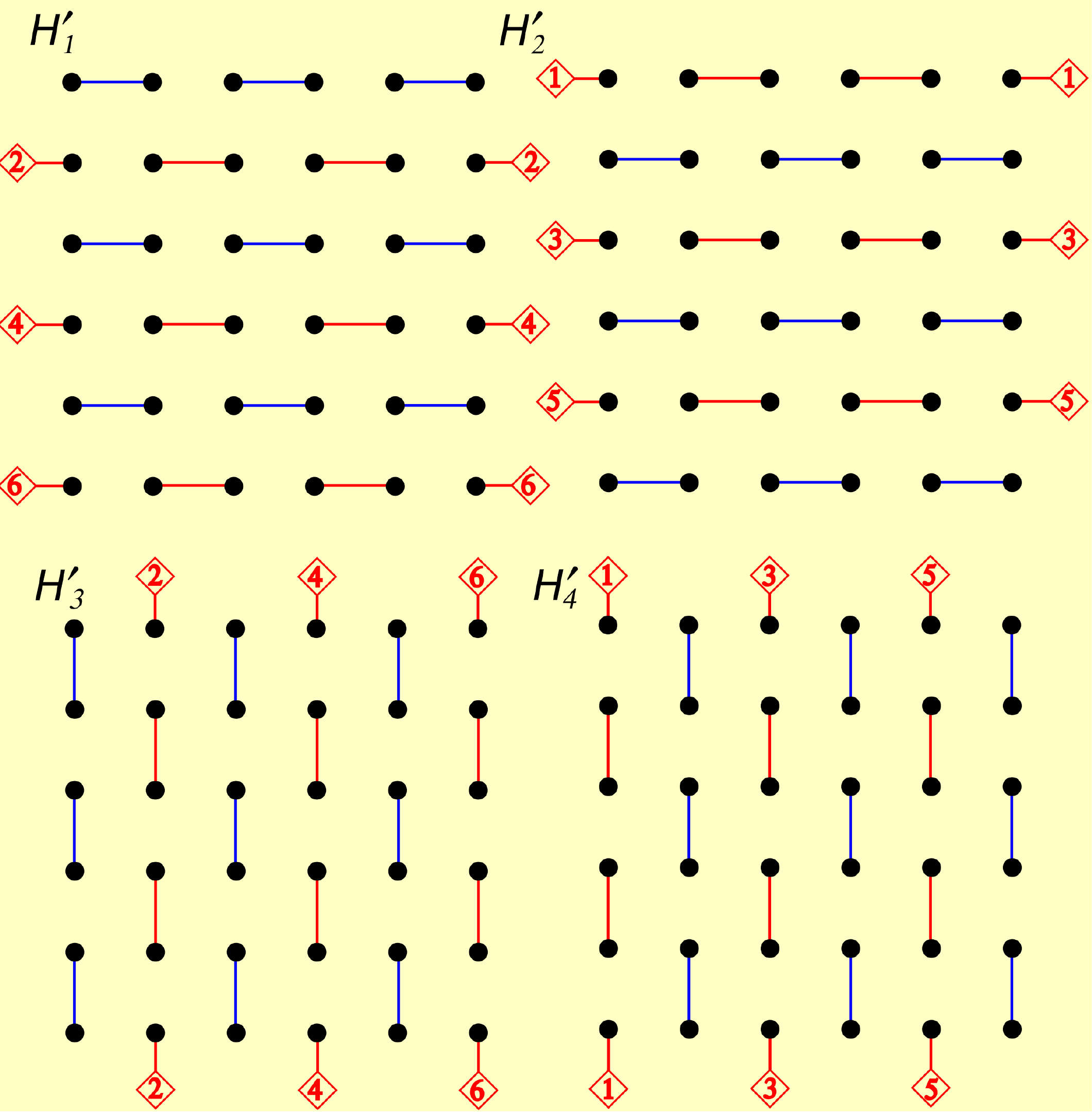}
\caption{Configuration of the turned-on couplings in the Hamiltonian (\ref{eq:tight-binding}) to realize the non-commutating SQW Hamiltonians (cf. Fig. \ref{fig:hamiltonians_naive}), with $d=6$. In each diagram the ending sites specified with the same numbers inside the diamonds are connected.}
\label{fig:hamiltonians}
\end{figure}

The Hamiltonians (\ref{eq:H_1_naive}) and (\ref{eq:H_2_naive}) commute with those in Eqs. (\ref{eq:H_3_naive}) and (\ref{eq:H_4_naive}), which implies the dynamics on the rows is independent of the dynamics on the columns. In fact, such Hamiltonians generate two 1D SQWs in the horizontal and vertical directions. One way to couple the dynamics in these two directions to obtain a 2D dynamics is to change the way the pairs of the coupled resonators are selected in the above constructions. For example, we can turn on those couplings in Eq. (\ref{eq:tight-binding}) that generate the Hamiltonian (\ref{eq:H_1D1}) for the odd rows and the Hamiltonian (\ref{eq:H_1D2}) for the even rows. The turned-on couplings in this case are shown in Fig. \ref{fig:hamiltonians} below $H'_1$. The Hamiltonian takes the form
\begin{equation}
\label{eq:H_1}
H'_1 =
\openone ^{10}_d \otimes H^{\mathrm{1D}}_1 +
\openone ^{01}_d \otimes H^{\mathrm{1D}}_2,
\end{equation}
where
$$\openone ^{10}_d = \openone _{d/2} \otimes
\begin{pmatrix}
       1&\;0\\
       0&\;0\\
\end{pmatrix} \;\;\;\;,\;\;\;
\openone ^{01}_d = \openone _{d/2} \otimes
\begin{pmatrix}
       0&\;0\\
       0&\;1\\
\end{pmatrix}.$$

The second Hamiltonian is constructed similarly but by interchanging the place of the 1D terms in the Hamiltonian (\ref{eq:H_1}). This is achieved by adjusting the couplings in Eq. (\ref{eq:tight-binding}) such that the Hamiltonians (\ref{eq:H_1D1}) and (\ref{eq:H_1D2}) are generated for the even and odd rows, respectively, and hence
\begin{equation}
\label{eq:H_2}
H'_2 =
\openone ^{10}_d \otimes H^{\mathrm{1D}}_2 +
\openone ^{01}_d \otimes H^{\mathrm{1D}}_1,
\end{equation}
which corresponds to~Fig. \ref{fig:hamiltonians} below $H'_2$.

The next two Hamiltonians are constructed using the same idea; however, the couplings in the Hamiltonian (\ref{eq:tight-binding}) are adjusted such that the 1D array Hamiltonians (\ref{eq:H_1D1}) and (\ref{eq:H_1D2}) are generated for the columns of the lattice, alternately. In Fig. \ref{fig:hamiltonians} below $H'_3$ and $H'_4$, the desired couplings in these cases are shown where the corresponding Hamiltonians can be written as
\begin{align}
\label{eq:H_3}
&H'_3 =
H^{\mathrm{1D}}_1 \otimes \openone ^{10}_d +
H^{\mathrm{1D}}_2 \otimes \openone ^{01}_d, \\
\label{eq:H_4}
&H'_4 =
H^{\mathrm{1D}}_2 \otimes \openone ^{10}_d +
H^{\mathrm{1D}}_1 \otimes \openone ^{01}_d. \\
\nonumber
\end{align}
We note that the Hamiltonians (\ref{eq:H_1}) and (\ref{eq:H_2}) do not commute with those in Eqs. (\ref{eq:H_3}) and (\ref{eq:H_4}); thus, these Hamiltonians generate a genuinely 2D SQW.

The Hamiltonian form (\ref{eq:tight-binding}) can be controlled such that the Hamiltonians (\ref{eq:H_1_naive}) - (\ref{eq:H_4_naive}) or (\ref{eq:H_1}) - (\ref{eq:H_4}) are generated in a sequence. Such sequence generates the SQW operator. For instance, the sequence $H'_4H'_2H'_3H'_1$, each for the period $\tau$, within the time interval $[0,4\tau)$, corresponds to the time evolution
\begin{equation}
\label{eq:evolution}
U(0,4\tau) = e^{-i\tau H'_4} e^{-i\tau H'_2}
                       e^{-i\tau H'_3} e^{-i\tau H'_1}.
\end{equation}
Calculating such evolution is reduced to obtaining the evolutions of the 1D Hamiltonians (\ref{eq:H_1D1}) and (\ref{eq:H_1D2}), which is achieved by noting
\begin{equation}
\label{eq:coin}
e^{ -i\tau ( \omega \openone _2 - \kappa \sigma_x ) } = e^{-i \omega \tau}
\begin{pmatrix}
\cos\kappa\tau & i \sin\kappa\tau \\
i \sin\kappa\tau & \cos\kappa\tau \\
\end{pmatrix}.
\end{equation}
Recalling that all resonators are in resonance and setting $\kappa\tau=2\pi \ell+\pi/4$, for an integer $\ell$, the evolution (\ref{eq:evolution}) implements the SQW dynamics that was used in Ref. \cite{portugal2017quantum} for designing an efficient quantum search algorithm.

To realize the SQW dynamics the system is initialized by generating a photon in one of the resonators, e.g., in the middle of the lattice,
\begin{equation}
\label{eq:initial_state}
|\psi_0\rangle = \big| (d^2- d)/2 \big\rangle.
\end{equation}
After applying the evolution (\ref{eq:evolution}) $l$ times, for the total time period $4\tau l$, the system evolves to the final state $U^l|\psi_0\rangle$.
The system can then be measured to find the photon in one of the resonators. For a system composed of superconducting microwave resonators coupled through \textsc{squid} elements, similar protocols as suggested in Ref. \cite{moqadam2017staggered} can be used to initialize and measure the
quantum walk. The probability distribution of finding the photon in the lattice is given by
\begin{equation}
\label{eq:prob_dist}
P_l(n) = \big| \langle n | U^l|\psi_0 \rangle \big|^2,
\end{equation}
where $\big\{|n\rangle;\,n=1,\ldots,N=d^2\big\}$ is the canonical basis for the lattice Hilbert space.

\section{Walk on 2D manifolds}
\label{sec:3}

We described how to control the couplings in the system Hamiltonian (\ref{eq:tight-binding}) in order to realize the SQW dynamics on a 2D lattice with the periodic boundary conditions in both directions. In this section, we explore the quantum walk dynamics on various 2D manifolds or surfaces which are obtained through modifying the boundary conditions.

A 2D manifold is a topological space that locally has the structure of the Euclidean plane $\mathbbmss{R}^2$. Basic 2D manifolds can be obtained by appropriately identifying the boundaries of a square \cite{nakahara2003geometry}. For instance, by identifying two opposite sides of a square, a cylinder is obtained; identification the other two sides creates a torus. For the 2D square lattice, identification of the boundaries can be performed by connecting the boundary sites by edges, i.e., by coupling any two sites that are supposed to be identified.

In this manner, by using the periodic boundary conditions for the 2D lattice, a torus is obtained. As it is implied by Figs. \ref{fig:hamiltonians_naive} (b) and \ref{fig:hamiltonians}, $-\kappa^{-1}  (\sum_{i=1}^4 H_i - 4 \omega \openone _{d^2} )$ and $-\kappa^{-1}  (\sum_{i=1}^4 H'_i - 4 \omega \openone _{d^2} )$ correspond to the adjacency matrices for the 2D lattice with the torus topology. However, depending on the way the SQW Hamiltonians are constructed, the choices of the boundary conditions may not affect all SQW Hamiltonians. This can be seen in Fig. \ref{fig:hamiltonians_naive} (b), where the Hamiltonians $H_1$ and $H_3$ do not feel the boundary conditions. The repeated application of the dynamics (\ref{eq:evolution}) on the initial state (\ref{eq:initial_state}) gives then the evolution of a quantum walker on a torus.

\begin{figure}
\includegraphics[scale=.35]{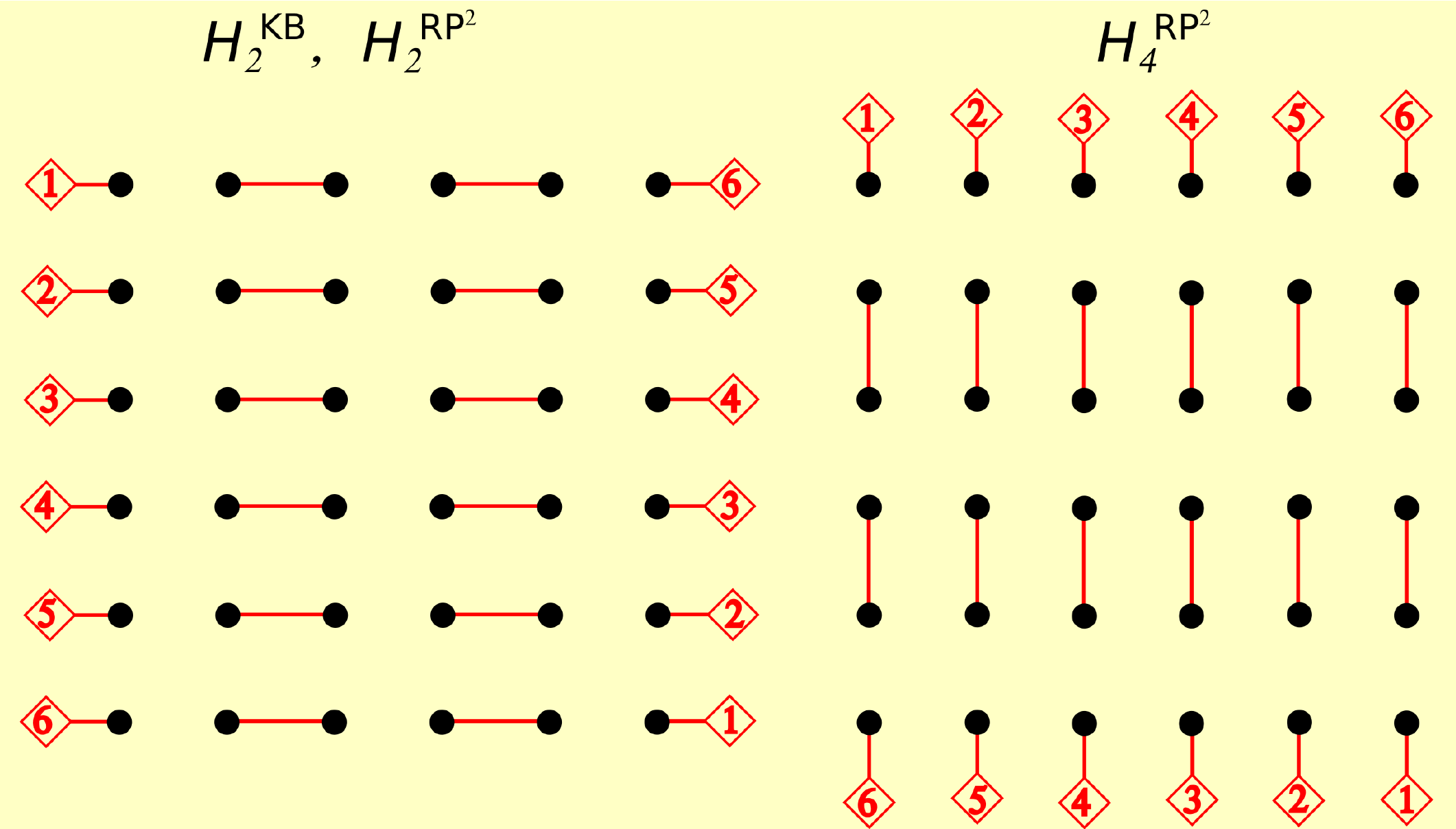}
\caption{Configuration of the turned-on couplings in the Hamiltonian (\ref{eq:tight-binding}) to realize the twisted boundary conditions, for $d=6$. In each diagram, the end sites specified with the same numbers inside the diamonds are connected.}
\label{fig:hamiltonians_KL_RP2}
\end{figure}

Another possibility for coupling two opposite sides of the 2D lattice is to twist one of the edges and then perform the identification. In this case, the sites on the boundaries are coupled in the opposite directions---Fig. \ref{fig:hamiltonians_KL_RP2}. The other two sides of the lattice can be coupled as
before. By doing such identification a 2D nonorientable surface, called Klein bottle, is obtained.
The SQW Hamiltonians generating the walk on the Klein bottle can be constructed similarly to the Hamiltonians (\ref{eq:H_1_naive}) - (\ref{eq:H_4_naive}) for the torus, depicted in Fig. \ref{fig:hamiltonians_naive} (b). Let us choose the Klein bottle boundary conditions in the horizontal direction, as seen in the left plot of Fig. \ref{fig:hamiltonians_KL_RP2}, and keep the vertical boundary conditions as before. Thus, the corresponding Hamiltonian $H^{\mathbbmss{KB}}_2$ takes a different form compared with $H_2$, given by
\begin{equation}
\label{eq:H_2_KB}
H^{\mathbbmss{KB}}_2 = \openone _d \otimes \widetilde{H}^{\mathrm{1D}}_2
                              -\kappa\, {\Sigma}_{\mathrm{hor}},
\end{equation}
where $\widetilde{H}^{\mathrm{1D}}_2$ is obtained from Eq. (\ref{eq:H_1D2}) by replacing two $-\kappa$s with $0$s in the first and the last rows, decoupling the first and the last resonators in each row of the lattice. The horizontal twisted boundary conditions are imposed by the matrix
\begin{equation}
{\Sigma}_{\mathrm{hor}} = J_d \otimes
                                                      \big( \sigma^+_d + \sigma^-_d \big),
\end{equation}
where $J_d$ is the row reversed version of the $d$-dimensional identity matrix (a matrix with $1$s on the main anti-diagonal and $0$s elsewhere)
and $\sigma^+_d$ ($\sigma^-_d$) is equal to the $d$-dimensional zero matrix except for the top-right (down-left) corner entry which is replaced with $1$.

The new boundary condition, however, does not modify $H^{\mathbbmss{KB}}_1$ with respect to $H_1$. Using the periodic boundary conditions for the
columns of the lattice, as before, the vertical SQW Hamiltonians for the Klein bottle ($H^{\mathbbmss{KB}}_3$ and $H^{\mathbbmss{KB}}_4$) remain equal to $H_3$ and $H_4$ (respectively).

If the twisted boundary conditions are applied for both directions (see Fig. \ref{fig:hamiltonians_KL_RP2}),
namely the opposite edges of the 2D lattice are twisted and then identified, the resulting 2D manifold is
the real projective plane.
The SQW dynamics on such a surface is generated by the Hamiltonians $H^{\mathbbmss{RP}^{2}}_i (i=1,2,3)$,
equal to the Hamiltonians for the walk on the Klein bottle, and
\begin{equation}
\label{eq:H_4_RP2}
H^{\mathbbmss{RP}^{2}}_4 = \widetilde{H}^{\mathrm{1D}}_2 \otimes \openone _d
                              -\kappa\, {\Sigma}_{\mathrm{ver}},
\end{equation}
where the vertical twisted boundary conditions are given by
\begin{equation}
{\Sigma}_{\mathrm{ver}} = \big( \sigma^+_d + \sigma^-_d \big) \otimes
                                                             J_d.
\end{equation}

\begin{figure}
\includegraphics[scale=.35]{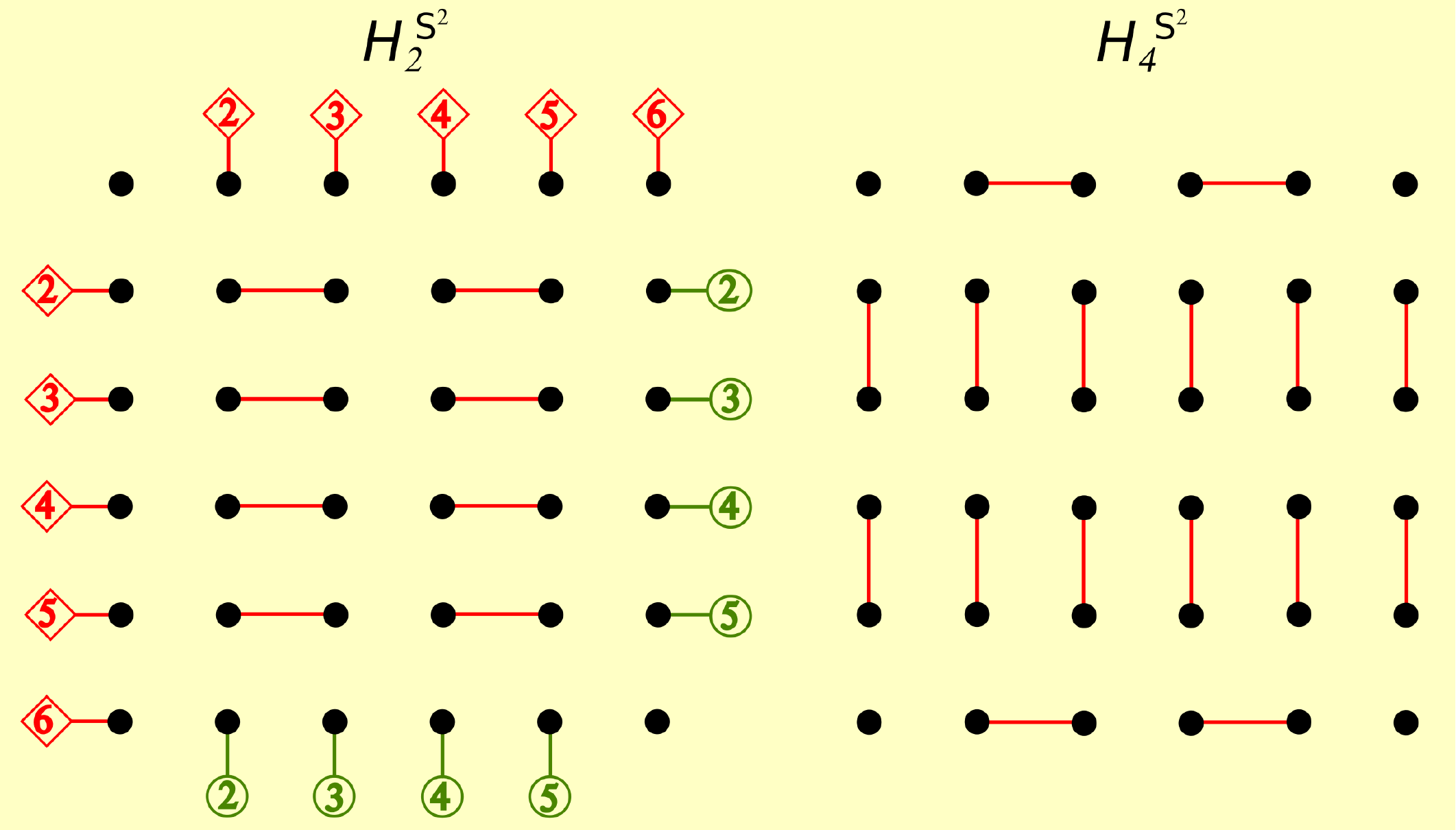}
\caption{Configuration of the turned-on couplings in the Hamiltonian (\ref{eq:tight-binding}) to realize the sphere boundary conditions, for $d=6$. In each diagram the ending sites specified with the same numbers inside the same boxes (diamonds or circles) are connected.}
\label{fig:hamiltonians_S2}
\end{figure}

Finally, by identification of the adjacent sides (rather than the opposite sides) of the 2D lattice a sphere is obtained. The staggered Hamiltonians for the walk on the sphere, $H^{\mathbbmss{S}^{2}}_i (i=1,2,3,4)$, can be constructed such that the first and the third Hamiltonians remain the same as $H_1$ and $H_3$, similarly to the previous cases. The turned-on coupling for the other two Hamiltonians are shown in Fig. \ref{fig:hamiltonians_S2}. The second Hamiltonian is given by
\begin{equation}
\label{eq:H_2_S2}
H^{\mathbbmss{S}^{2}}_2 = 
\begin{pmatrix}
\omega \openone _d & -\kappa P_1                                & -\kappa P_2 \\
-\kappa P^T_1 & \openone _{d-2} \otimes \widetilde{H}^{\mathrm{1D}}_2 & -\kappa Q^T\\
-\kappa P^T_2                     & -\kappa Q             & \omega \openone _d \\
\end{pmatrix},
\end{equation}
which is the sum of a block diagonal matrix, obtained from the Hamiltonian (\ref{eq:H_2_naive}) by decoupling the sites on all four sides of the lattice, and the desired boundary conditions, shown in the left diagram in Fig. \ref{fig:hamiltonians_S2}. In this equation, the boundary conditions are incorporated to
\begin{align*}
P_{d \times d^2-d} &= ( \mathbf{0}_{d-1\times 1} \quad \openone _{d-1})^{T}
                                 \otimes ( 1 \quad \mathbf{0}_{1\times d-1}), \\
Q_{d \times d^2-2d} &= ( \mathbf{0}_{d-2\times 1} \quad \openone _{d-2}
                                  \quad \mathbf{0}_{d-2\times 1} )^{T}
                                  \otimes ( \mathbf{0}_{1\times d-1} \quad 1).
\end{align*}
where we have represented $P$ by $[P_1 \;\; P_2]$ and ``$T$" denotes transposition. The fourth Hamiltonian (see the right diagram in Fig. \ref{fig:hamiltonians_S2}) takes the form
\begin{equation}
\label{eq:H_4_S2}
H^{\mathbbmss{S}^{2}}_4 = 
\begin{pmatrix}
\widetilde{H}^{\mathrm{1D}}_2 &                                       &               \\
             & H^{\mathrm{1D}_{(d-2)}}_1 \otimes \openone _{d}  &               \\
             &                                       & \widetilde{H}^{\mathrm{1D}}_2  \\
\end{pmatrix},
\end{equation}
where $H^{\mathrm{1D}_{(d-2)}}_1$ is similar to the 1D Hamiltonian (\ref{eq:H_1D1}) but for a chain of $d-2$ sites, and all other elements are $0$.
Note that the adjacency matrix of the lattice with the sphere topology is given by $-\kappa^{-1}  (\sum_i H^{\mathbbmss{S}^{2}}_i - 4 \omega \openone _{d^2} )$, which includes $2d^2-3$ edges---$3$ edges less than the previous cases.

For the Klein bottle, the real projective plane, and the sphere, the more complex version of the SQW Hamiltonians, based on Fig. \ref{fig:hamiltonians}, can also be constructed. However, since our objective in this paper is to study effects of the boundary conditions, we skip those cases.

We remark that among the surfaces considered in this section, the sphere and the torus are orientable surfaces with genus $0$ and $1$, respectively. In contrast, the real projective plane and the Klein bottle are non-orientable surfaces with non-orientable genus equal to $1$ and $2$, respectively.

\section{Boundary-induced coherence}
\label{sec:4}

Having considered a 2D manifold and the corresponding staggered Hamiltonians, as described in the previous section, the SQW dynamics $U$ can be constructed according to Eq. (\ref{eq:evolution}). After applying the $U$ operator $l$ times on the localized initial state of the walker (photon) [Eq. (\ref{eq:initial_state})], the density matrix of the photon is obtained as
\begin{equation}
\varrho_{l} = U^l |\psi_0 \rangle \langle \psi_0 | U^l.
\end{equation}
The coherence of the (pure) state of the photon at a given step $l$ can be quantified by \cite{baumgratz2014quantifying}
\begin{equation}
\label{eq:coherence}
C_l = \sum_{n,m=1}^N \big|[\varrho_{l}]_{nm}\big|-1.
\end{equation}
The diagonal elements of the density matrix ($[\varrho_{l}]_{nn}$) give the probability distribution of finding the photon in
different sites, as computed by Eq. (\ref{eq:prob_dist}).
The Shannon entropy of the system at step $l$ is then calculated by
\begin{equation}
\label{eq:entropy}
E_l = - \sum_{n=1}^{N} [\varrho_{l}]_{nn}\log_2 [\varrho_{l}]_{nn}.
\end{equation}

The coherence and the entropy are appropriate quantities for addressing quantumness in behavior of the quantum walk on different topologies \cite{sarovar2010quantum,baumgratz2014quantifying,shahbeigi2018quantum}. In fact, the von Neumann entropy cannot be used here, since it is identically $0$ for the isolated system.

\begin{figure}
\includegraphics[scale=.38]{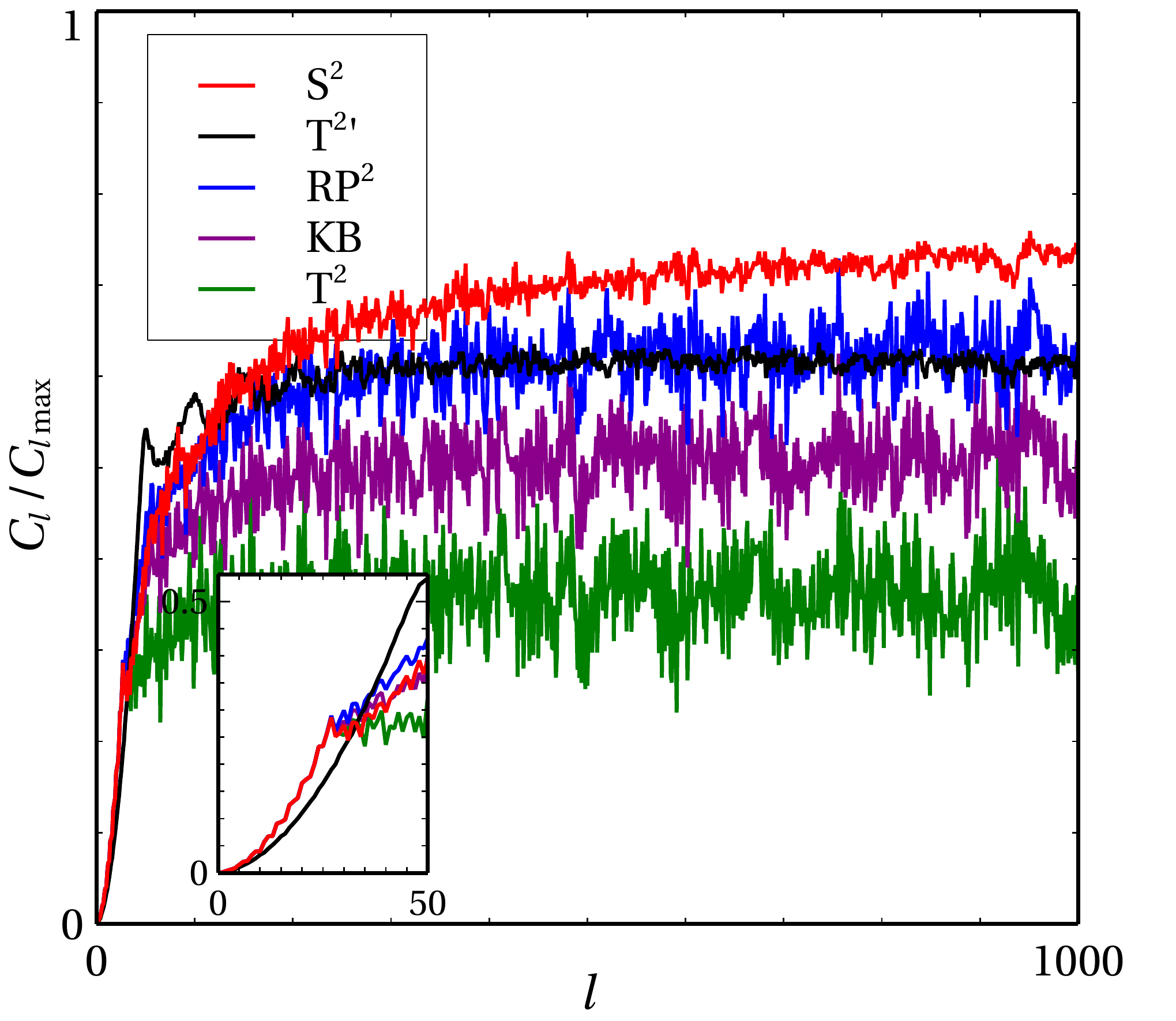}
\caption{The normalized coherence [Eq. (\ref{eq:coherence})] for a 2D lattice with different boundary conditions. The plots are generated by using the state (\ref{eq:initial_state}) as the initial state of the walker and setting $d=100$, $n=1000$, $\kappa\tau= \pi/3$, and $\omega\tau=2\pi$. The inset shows the coherence for the first $50$ steps.}
\label{fig:coherence}
\end{figure}

Figure \ref{fig:coherence} shows $C_l/{C_l}_\mathrm{max}$ in terms of the number of steps for quantum walks on the torus with the staggered Hamiltonians (\ref{eq:H_1_naive}) - (\ref{eq:H_4_naive}), the Klein bottle with $H_i^{\mathbbmss{KB}}$, the real projective plane with $H_i^{\mathbbmss{RP}^{2}}$, the torus with the Hamiltonians (\ref{eq:H_1}) - (\ref{eq:H_4}), and the sphere with $H_i^{\mathbbmss{S}^{2}}$. The corresponding dynamics are labeled by $\mathbbmss{T}^{2}$, $\mathbbmss{KB}$, $\mathbbmss{RP}^{2}$, ${\mathbbmss{T}^{2}}'$, and
$\mathbbmss{S}^{2}$, respectively. In generating those plots, the system frequencies are set such that $\kappa\tau= 2\pi\ell + \pi/3$ and
$\omega\tau=2\pi\ell'$, for some integers $\ell$ and $\ell'$. The lattice size is fixed to $N=100\times100$ sites.

The coherence takes its minimum ${C_l}_{\mathrm{min}}=0$ for the initial localized state given in Eq. (\ref{eq:initial_state}). It increases, then, for all the boundary conditions until the photon-wave function populates the boundary sites. It can be seen that within that interval the envelope function of the coherence is a convex function (see the inset in Fig. \ref{fig:coherence}) but later it changes to a concave function. Except for the case ${\mathbbmss{T}^{2}}'$, the other dynamics are indistinguishable until the boundaries are met. A single populated site can affect up to the second neighbors, at each step of the staggered evolution. Hence the effects of the boundary conditions, for the photon initially located at the center, appears
around the step $l=d/4$. For the case ${\mathbbmss{T}^{2}}'$, due to the interference, it takes more steps until the boundaries are sufficiently populated and the interference between different parts of the wave-function is started.

The walk on the torus (green plot labeled with ${\mathbbmss{T}^{2}}$) has the lowest level of coherence during the whole dynamics. This can be justified by recalling the staggered Hamiltonians corresponding to this case generate two \textit{independent} 1D quantum walks. The coherence increases by applying twisted boundary conditions which make the horizontal and vertical dynamics correlated. The plots of $\mathbbmss{KB}$ and $\mathbbmss{RP}^{2}$ lie above the plot of $\mathbbmss{T}^{2}$. The dynamics on $\mathbbmss{S}^{2}$ has the highest coherence. The 2D quantum walk on the torus (black plot labeled with ${\mathbbmss{T}^{2}}'$) generates the coherence comparable with the coherence for the walk on the real projective plane ($\mathbbmss{RP}^{2}$). The oscillatory behavior of the coherence plots is considerably decreased for the dynamics corresponding to
the case ${\mathbbmss{T}^{2}}'$. 

\begin{figure}[tp]
\includegraphics[scale=.38]{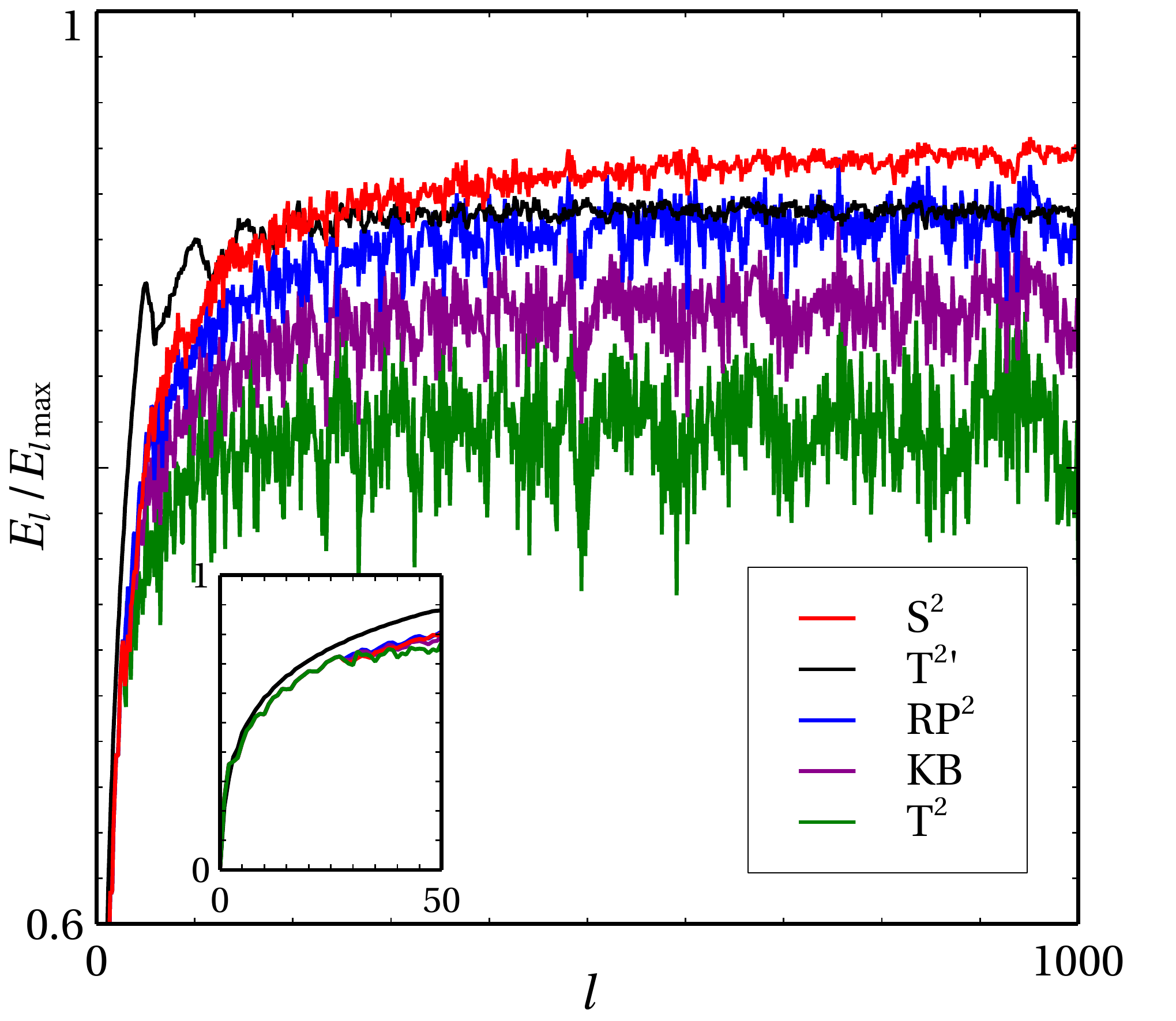}
\caption{The normalized entropy [Eq. (\ref{eq:entropy})] for a 2D lattice with different boundary conditions. The plots are generated by using (\ref{eq:initial_state}) as the initial state of the walker and setting $d=100$, $n=1000$, $\kappa\tau= \pi/3$, and $\omega\tau=2\pi$. The inset shows the entropy for the first $50$ steps.}
\label{fig:entropy}
\end{figure}
The maximum of the coherence, corresponds to the diagonal state, the maximally coherent state \cite{baumgratz2014quantifying}
\begin{equation}
\label{eq:maximally_coherent}
| \psi_{\mathrm{diag}} \rangle = \sum_{n=1}^N | n \rangle/\sqrt{N},
\end{equation}
and its value is ${C_l}_{\max}=d^2-1$.
The plots in Fig. \ref{fig:coherence} are far from the maximum value ${C_l}/{C_l}_{\max}=1$ associated with the state (\ref{eq:maximally_coherent}) and apparently the state of the photon never converges to that diagonal state.

Figure \ref{fig:entropy} shows $E_l/{E_l}_{\max}$ for quantum walks on a lattice with different boundary conditions. The entropy is $0$ for the localized initial state (\ref{eq:initial_state}) and then increases having a concave envelope function in terms of the walk steps (see also the inset in Fig. \ref{fig:entropy}). The walk on the torus (green plot labeled with $\mathbbmss{T}^{2}$) has the lowest
level for the entropy. The next levels correspond to the walk on the Klein bottle and the real projective plane, consecutively, which have twisted boundary conditions. The entropy of the walk on the torus labeled with ${\mathbbmss{T}^{2}}'$ is about the same level as the walk on the real projective plane.
The walk on the sphere generates the largest level for the entropy. The maximum value of the entropy, ${E_l}_{\max} = 2\log_2 d$, seems not to be achieved by the SQW dynamics. In this manner, the entropy (related to the the diagonal elements of the density matrix) represents qualitatively a similar information as the coherence implies.

It is seen that the choice of the boundary conditions affects the quantum walk dynamics by modifying the interference pattern. The dynamics of the coherence and the entropy are distinct for quantum walks on different manifolds. In particular, the coherence associated with the 2D SQW on the torus can be reached by generating independent 1D SQWs in two dimensions but modifying the boundary conditions. This implies that, the coherence is indeed induced by the boundaries.

The observation that the dynamics of the coherence and the entropy qualitatively represent similar information suggests that the entropy can also quantify the coherence in quantum walks. In the following sections, we investigate the dynamics of the entropy for the CTQW on different topologies. This analysis gives an indication of the boundary-induced coherence in the walk. We then explore the dynamics of the entropy for the random walk to see the boundary effects in the absence of the quantum coherence.

\section{CTQW}
\label{sec:5}

The SQW formalism can be employed to approximate CTQW dynamics. The approximation error is given by the generalized decomposition relation \cite{suzuki1985decomposition,suzuki1976generalized}
\begin{align}
\label{eq:suzuki_theorem}
\biggl \| e^{-\sum_{j=1}^{4} it H_j} -
\biggl[ \prod_{j=1}^{4} e^{-i\frac{t}{L} H_j} \biggr]^L \biggr\| \leqslant \; \frac{t^2}{2L} \sum_{j>k} \big\Vert  [ H_j,H_k ] \big\Vert ,
\end{align}
where $\{H_j; j=1,2,3,4\}$ is any set of the staggered Hamiltonians discussed in the previous sections, $t$ is a given period of time, $L$ is an integer, and $\Vert X\Vert =\sup_{\Vert v\Vert=1}|\langle v|X|v\rangle|$ (with the Euclidean vector norm $\Vert v\Vert=\sqrt{\langle v|v\rangle}$) is the standard operator norm. The left-hand side of the inequality corresponds to the difference between a CTQW evolving for the total time $t$ and an $L$-step SQW dynamics. The difference, as given in the right-hand side of the inequality, is bounded and can be decreased by increasing $L$. Note that the diagonal elements of the CTQW Hamiltonian simulated by the SQW generators are $4$ times larger than the diagonal elements (resonator frequencies) that appear in the direct construction of the CTQW dynamics. However, since it is supposed that all the the resonators are in resonance, this modification has no effect on the dynamics.  

\begin{figure}
\includegraphics[scale=.38]{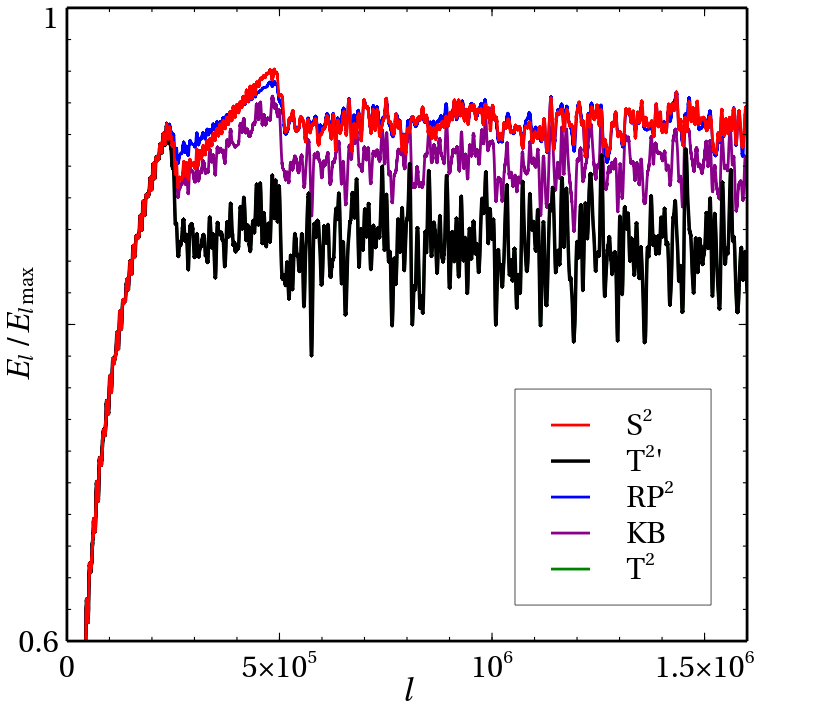}
\caption{The normalized entropy [Eq. (\ref{eq:entropy})] in terms of the number of steps approximated for the CTQW on a 2D lattice with different boundary conditions. The plots are generated by using an SQW with the initial state (\ref{eq:initial_state}) and setting $d=100$, $n=16\times10^5$, $\kappa\tau= 10^{-4}$, and $\omega\tau=2\pi$.}
\label{fig:entropy_ctqw}
\end{figure}

To obtain an upper bound in Eq. (\ref{eq:suzuki_theorem}), first we write the staggered Hamiltonians as $ H_j = \omega\, \openone _{d^2} - \kappa\, G_j$, in order to separate the contribution of the resonator and the coupling frequencies. In fact, $G_j$ are the adjacency matrices of the lattices associated with $H_j$. In this manner, the terms containing the resonator frequency are canceled out in the commutator brackets
\begin{align}
\Vert  [ H_j,H_k ] \Vert  \leqslant 
                  2\kappa^2 \; \Vert  G_j \Vert  \; \Vert  G_k \Vert.
\end{align}
The matrices $G_j$ are orthogonal reflections \citep{portugal2016staggered-qip}, namely Hermitian and unitary. This yields an upper bound error as
\begin{equation}
\varepsilon \leqslant 6 \kappa^{2} t^{2}/L.
\end{equation}

As an example, to simulate a CTQW dynamics on a lattice with the fixed coupling frequency $\kappa=1$ MHz and for the total time evolution $t=(1/6)\times\!10^{-3}$ s, we can set the total number of steps for the SQW to $L\approx\!(1/6)\times 10^{7}$. This leads to $\kappa \tau = \kappa t/L = 10^{-4}$ and the error is bounded by $10\%$. The total time for realizing the SQW dynamics, however, is $4t$. The total time here is comparable with the total time considered in the previous section for realizing $10^{3}$ steps of the SQW dynamics with $\kappa \tau = \pi/3$, namely $4.18\times 10^{3}$ s.

Figure \ref{fig:entropy_ctqw} shows the approximated normalized entropy for the CTQW, using the SQW dynamics. The CTQW dynamics on the tori correspond to the summations $\sum_{i=1}^{4}H_i$ and $\sum_{i=1}^{4}H'_i$ which are identical; thus, the plots related to the cases $\mathbbmss{T}^{2}$ and ${\mathbbmss{T}^{2}}'$ coincide in this figure. It is also seen that the plots associated with the walk on the tori lie lower than the plots with the twisted boundary conditions, and in this respect there is no qualitative difference between the CTQW and the SQW dynamics. The dynamics associated with the sphere (red plot labeled with $\mathbbmss{S}^{2}$) almost coincides with the plot for the Klein bottle (blue plot labeled with $\mathbbmss{KB}$), for the CTQW. Moreover, the effects of the boundary conditions appear around step $l = 10^4(d/2)$, when the boundary sites are sufficiently populated.

The entropy dynamics are then discernible for the CTQW evolutions on different 2D manifolds. In fact, the CTQW dynamics can reveal the topology of the underlying surface. It should be remarked that the direct calculation of the CTQW dynamics on a lattice of the size $N=100\times 100$ requires obtaining the evolution of Hamiltonians of the size $10^4\times10^4$ the computational cost of which is relatively high. However, the above approximation provides a means to calculate the CTQW dynamics with significantly less computational resources. Of course, it is still relatively costly to calculate the coherence for $16\times10^5$ steps of the SQW, and hence we have resorted to calculate the entropy. The above analysis, however, indicates that the boundary-induced coherence can be reflected by the dynamics of the entropy too.

\section{Random walk}
\label{sec:6}

To compare the quantum walk dynamics with the classical random walk behavior, we analyze the random walk evolution on the 2D manifolds. The desired classical dynamics can be generated by modifying the staggered evolutions. The SQW operators are block-diagonal and the blocks are given by
Eq. (\ref{eq:coin}). Substituting each block with
\begin{equation}
\label{eq:coin_classic}
\begin{pmatrix}
\cos^2\kappa\tau & \sin^2\kappa\tau \\
\sin^2\kappa\tau & \cos^2\kappa\tau \\
\end{pmatrix},
\end{equation}
we obtain different sets of doubly-stochastic matrices $\{ U^{\mathrm{cl}}_i ; i=1,2,3,4 \}$ corresponding to different boundary conditions. The discrete-time 2D random walk dynamics can be generated by applying the doubly-stochastic matrix
\begin{equation}
U^{\mathrm{cl}} = \frac{1}{4} \sum_{i=1}^4 U^{\mathrm{cl}}_i,
\end{equation}
on the (classical) initial state (\ref{eq:initial_state}). Indeed, the diagonal elements are the probabilities that the walker stays at each site and the off diagonal terms are the probabilities that the walker jumps to the neighboring sites.

\begin{figure}[tp]
\includegraphics[scale=.38]{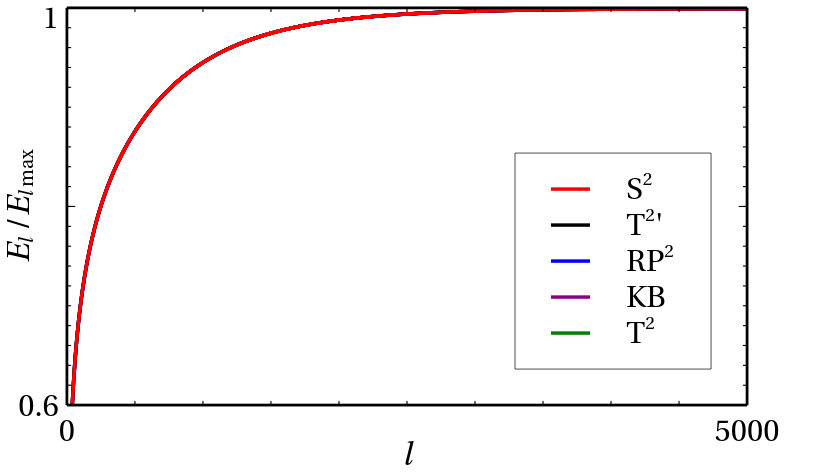}
\caption{The normalized entropy [Eq. (\ref{eq:entropy})] in terms of the number of steps for the random walk on a 2D lattice with different boundary conditions. The plots are generated by using (\ref{eq:initial_state}) as the initial state of the walker and setting $d=100$, $n=16\times10^3$, and $\kappa\tau= \pi/3$ in Eq. (\ref{eq:coin_classic}). All plots coincide.}
\label{fig:entropy_rw}
\end{figure}

Figure \ref{fig:entropy_rw} shows the normalized entropy for the random walks on different 2D manifolds discussed in the previous sections. It can be seen that the entropy behavior is independent of the choice of boundary conditions. Comparing Fig. \ref{fig:entropy_rw} with Figs. \ref{fig:entropy}~and \ref{fig:entropy_ctqw} reveals a sharp difference between the classical and the quantum dynamics on 2D manifolds. The interference causes the quantum walk dynamics to be sensitive to the boundary conditions which is manifested in the entropy evolution. However, for the random walk, there is no interference and the entropy dynamics is identical for all the manifolds. In fact, the insensitivity of the entropy dynamics to the boundary conditions in random walks supports that the different levels of the entropy value in quantum walks can reflect the quantumness of the walks modified by the boundaries. Note that, as expected, the maximum value of the entropy ${E_l}_{\mathrm{max}} = 2\log_2 d$ is achieved for the random walk, in finite time steps.

\vspace{0.2cm}

\section{Summary and discussions}
\label{sec:7}

We have designed the required Hamiltonians for the realization of quantum walks on 2D manifolds. The coinless discrete-time SQW model has been considered, which can be implemented on a lattice of superconducting microwave resonators interacting with tunable couplings.
Using periodic boundary conditions, we have devised two sets of SQW Hamiltonians. One set generates two uncoupled 1D walks and the other set corresponds to a 2D quantum walk. By changing the lattice boundary conditions, surfaces with different topologies can be obtained. We have given the explicit forms of the SQW Hamiltonians for quantum walk on various surfaces and investigated the properties of
the corresponding dynamics.

We have also explored the coherence and entropy for the walk on some specific 2D manifolds, such as torus, Klein bottle, real projective plane and sphere. We have shown that both functions are sensitive to the boundary conditions having distinct, discernible behaviors for the walk on different manifolds. It has been shown that these quantities have larger average values for lattices with twisted boundary conditions.

We have also considered the behavior of the CTQW and the corresponding classical random walk on the 2D manifolds and compared the results with the SQW dynamics. It has been observed that whereas the entropy is resolved for the quantum walks on different manifolds, it takes a fixed value for the classical random walk on all of those 2D surfaces.

For the analysis of the SQW in this paper, we fixed the initial place of the walker to the center of the lattice, the frequency of each step to $\kappa\tau=\pi/3$, and the size of the lattice to $N=100\times100$. Further numerical simulations (not reported here) have indicated that the general picture is fairly stable versus variations in the frequency and some translations of the initial state. Increasing the size of the lattice will decrease the amplitude of the oscillations in the coherence and the entropy, but changing the order of the SQW Hamiltonians does not change the general picture.

The dynamics explored in this paper can be used to simulate topological insulators in two dimensions. It may provide a tool to investigate the electron dynamics on Fermi surfaces, which are 2D manifolds embedded in the Brillouin zone of a crystal. Moreover, it may provide a way to study the efficiency of quantum-walk-based algorithms on databases with topological structures.

\textit{Acknowledgments.}---This work was supported by the Iran National Elites Foundation under Grant No. 7000/2000-1396/03/08 (to J.K.M.) and Sharif University of Technology's Office of Vice President for Research (to A.T.R.).

%

\end{document}